\documentclass[11pt]{article}
\usepackage{times}
\usepackage{mathfont}
\usepackage{graphicx}
\usepackage{subfigure}

\def\log{\mathop{{\rm log}}}
\let\mathbf\rm   

\begin{document}

\title{The Distribution of
Cycle Lengths in Graphical Models for Iterative Decoding}

\author{
Xian-ping Ge, David Eppstein, Padhraic Smyth \\
Information and Computer Science \\
University of California at Irvine \\
Irvine, CA 92697-3425 \\
{\tt \{xge,eppstein,smyth\}@ics.uci.edu}}

\maketitle

\begin{abstract}

This paper analyzes the distribution of cycle lengths in
turbo decoding and low-density parity  check (LDPC) graphs.  The
properties of such cycles are of significant interest
in the context of iterative decoding algorithms which are based
on belief propagation or message passing. We estimate the
probability that there exist no simple cycles of length less than
or equal to $k$ at a randomly chosen node in a turbo decoding
graph using a combination of counting arguments and independence
assumptions. For large block lengths $n$, this probability is
approximately $e^{-\frac{2^{k-1}-4}{n}}, k \ge 4$.
 Simulation results validate the accuracy of the
various approximations.  For example, for turbo codes with a block
length of 64000, a randomly chosen node has a less than $1\%$
chance of being on a cycle of length less than or equal to 10, but
has a greater than $99.9\%$ chance of being on a cycle of length
less than or equal to 20. The effect of the ``S-random"
permutation is also analyzed and it is shown that while it
eliminates short cycles of length $k < 8$, it does not
significantly affect the overall distribution of cycle lengths.
Similar analyses and simulations are also presented for graphs for
LDPC codes. The paper concludes by commenting
briefly on how these results may provide insight
into the practical success of iterative decoding methods.

\end{abstract}

\section{Introduction}
\label{sec:1.1}

Turbo codes are a new class of coding systems
that offer near optimal coding performance while requiring only
moderate decoding complexity [1].
It is known that the widely-used
iterative decoding algorithm for turbo codes is in
fact a special case of a  quite general
local message-passing algorithm for efficiently computing
posterior probabilities in acyclic directed graphical
(ADG) models (also
known as ``belief networks") [2, 3]. Thus, it is
appropriate to analyze the properties of iterative-decoding
by analyzing the properties of the associated ADG model.

In this paper we derive analytic approximations
for the probability that a randomly chosen node in
the graph for a turbo code participates in a simple
cycle of length less than or equal to $k$. The
resulting expressions provide insight into the
distribution of cycle lengths in turbo decoding.
For example, for block lengths of $64000$,
a randomly chosen node in the graph participates
in cycles of length less than or equal to
$8$ with probability 0.002, but participates
in cycles of length less than or equal to 20
with probability 0.9998.

In Section \ref{sec:notation} we review
briefly the idea of ADG models, define the
notion of a {\it turbo graph} (and the related
concept of a {\it picture}), and discuss how
the cycle-counting problem can be addressed
by analyzing how pictures can be embedded
in a turbo graph. With these basic
tools we proceed in Section \ref{sec:count}
to obtain closed-form expressions for
the number of pictures
of different lengths.  In Section \ref{sec:prob} we
derive upper and lower bounds on the
probability of embedding a picture in
a turbo graph at a randomly chosen node.
Using these results, in Section \ref{sec:nocycles}
we derive approximate expressions for the
probability of no simple cycles of length $k$
or less. Section \ref{sec:simulation} shows that
the derived analytical expressions are in
close agreement with simulation. In Section \ref{sec:srandom}
we investigate the effect of the S-random permuter
construction.  Section \ref{sec:ldpc} extends the
analysis to LDPC codes and compares both
analytic and simulation results on cycle lengths.
Section \ref{sec:discussion}
contains a discussion of what these results
may imply for iterative decoding in a general context and
Section \ref{sec:conclusion} contains the final conclusions.

\begin{figure}
\begin{center}
\leavevmode
\includegraphics{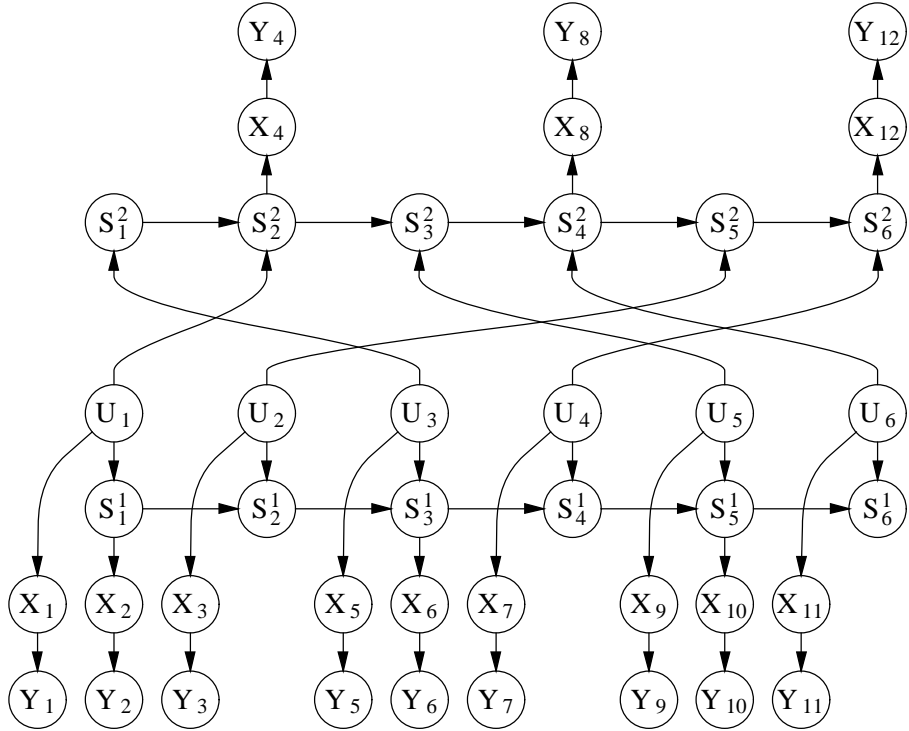}
\end{center}
\caption{The ADG model for a $K=6, N=12$, rate $1/2$
turbocode.}
\label{turbo1}
\end{figure}

\section{Background and Notation}
\label{sec:notation}

\subsection{Graphical Models for Turbo-codes}
An ADG model (also known as a belief network)
consists of a both a directed
graph and an associated probability distribution over
a set of random variables of interest. \footnote{Note that
``ADG" is the more widely
used terminology  in the statistical
literature, whereas the term belief network or
Bayes network is more widely used in computer science;
however, both frameworks are completely equivalent.}
There is a 1-1 mapping between the nodes in the
graph and the random variables. Loosely speaking,
the presence of a directed edge from
node $A$ to $B$ in the graph means that $B$ is
assumed to have a direct dependence on $A$ (`` $A$
causes $B$"). More generally, if we identify $\pi(A)$
as the set of all {\it parents} of $A$ in the graph (namely,
nodes which point to $A$), then
$A$ is conditionally independent of all other variables (nodes)
in the graph (except for $A$'s descendants) given the values
of the variables (nodes)
in the set $\pi(A)$. For example, a Markov chain is a special case
of such a graph, where each variable has a single parent.
The general ADG model framework is quite powerful in
that it allows us to systematically model and analyze
independence relations among relatively
large and complex sets of random variables [4].

As shown in [2, 3, 5], turbo codes
can be usefully cast in an ADG framework.
Figure \ref{turbo1} shows the ADG model
for a rate $1/2$ turbo code.
The $\mathbf{U}$ nodes
are the original information bits to be coded,
the $\mathbf{S}$ nodes are the
linear feedback shift register outputs,
the $\mathbf{X}$ nodes are the codeword vector which is the input
to the communication channel,
and the $\mathbf{Y}$ nodes are the channel outputs. The
ADG model captures precisely the conditional independence
relations which are implicitly assumed in the turbo coding
framework, i.e., the input bits $\mathbf{U}$ are
marginally independent,  the
state nodes $\mathbf{S}$ only depend on the previous state
and the current input bit, and so forth.

The second component of an ADG model (in addition to the
graph structure) is the specification of a joint probability
distribution on the random variables. A fundamental
aspect of ADG models is the fact that this joint
probability distribution decomposes into a simple factored
form. Letting $\{A_1,\ldots,A_n\}$ be the variables of interest, we have
\begin{equation}
p(A_1,\ldots,A_n) = \prod_{i=1}^n p\left(A_i|\pi(A_i)\right),
\end{equation}
i.e., the overall joint distribution is the product of the conditional
distributions of each variable $A_i$ given its parents $\pi(A_i)$.
(We implicitly assume discrete-valued variables here and refer to
distributions; however, we can do the same factorization with
density functions for real-valued
variables, or with combinations of densities and distributions).

To specify the full joint distribution, it is sufficient to
specify the individual conditional distributions. Thus, if
the graph is sparse (few parents) there can be considerable
savings in parameterization of the model. From a decoding viewpoint,
however, the fundamental advantage of this factorization
is that it permits the efficient
calculation of posterior probabilities (or
optimal bit decisions) of interest. Specifically,
if the values for a subset of variables are known (e.g.,
the received codeword vector $\mathbf{Y}$) we can
efficiently compute the posterior
probability for  the
information bits $\mathbf{U_i}=1$, $1 \le i \le N$. The
power of the ADG framework is that there exist exact
 local message-passing algorithms which
calculate such posterior probabilities. These
algorithms typically have time complexity which is linear in
the diameter of the underlying graph times
a  factor which is exponential in
the cardinality of the variables at the nodes
in the graph.   The algorithm is provably convergent to
the true posterior probabilities provided the
graph structure does not contain any loops (a loop
is defined as a cycle in the undirected version of
the ADG, i.e., the graph where directionality of the
edges is dropped). The
message-passing algorithm of Pearl [6] was the
earliest general algorithm (and is perhaps the
best-known) in this general class of ``probability
propagation" algorithms.  For
regular convolutional codes, Pearl's message passing
algorithm applied to the convolutional code
graph structure (e.g., the lower half of Figure 1)
directly yields the BCJR decoding algorithm [7].

If the graph has loops then
Pearl's algorithm no longer provably converges, with the
exception of certain special cases (e.g., see [8]). A ``loop"
is any cycle in the graph, ignoring directionality of the edges.
The turbocode ADG of Figure 1 is an example
of a graph with loops.
In essence, the messages being passed can arrive at the same
node via multiple paths, leading to multiple ``over-counting"
of the same information.

A widely used strategy in statistics and artificial intelligence
is to reduce the original graph with loops to an equivalent graph
without loops (this can be achieved by clustering variables
in a judicious manner) and then applying Pearl's algorithm to
the new graph. However, if one applies this method to ADGs for
realistic turbo codes the resulting graph (without loops) will
contain at least one node with a
large number of variables.
This node will have cardinality exponential in this number
of variables, leading
to exponential complexity in the probability
calculations referred to above. In the worst-case all
variables are combined into a single node and there is
in effect no factorization. Thus, for turbo codes, there
is no known efficient exact algorithm for computing posterior
probabilities (i.e., for decoding).

Curiously, as shown in [2, 3, 4], the iterative decoding algorithm
of [1] can be shown to be equivalent to applying the local-message
passing algorithm of Pearl directly to the ADG structure for
turbo codes (e.g., Figure 1), i.e., applying the iterative
message-passing algorithm to a graph with loops. It is well-known that in
practice this decoding strategy performs very well in
terms of producing lower bit error rates than any virtually other
current coding system of comparable complexity.
Conversely, it
is also  well-known that message-passing in graphs with
loops can converge to incorrect posterior probabilities (e.g., [9]).
Thus, we have the ``mystery" of turbo decoding: why does
a provably incorrect algorithm produce an extremely useful
and practical decoding algorithm? In the remainder
of this paper we take a step in understanding message-passing in graphs
with loops by characterizing the distribution of cycle-lengths
as a function of cycle length. The
motivation is as follows: if it turns out that cycle-lengths
are ``long enough" then there may be a well-founded basis
for believing that message-passing in graphs with cycles
of the appropriate length are not susceptible to
the ``over-counting" problem mentioned earlier (i.e., that
the effect of long loops in practice may be negligible).
This is somewhat speculative and we will return
to this point in  Section \ref{sec:discussion}. An
additional motivating factor is that the
characterization of cycle-length distributions in turbo codes
is of fundamental interest by itself.

\subsection{Turbo Graphs}

\begin{figure}
\begin{center}
\leavevmode
\includegraphics{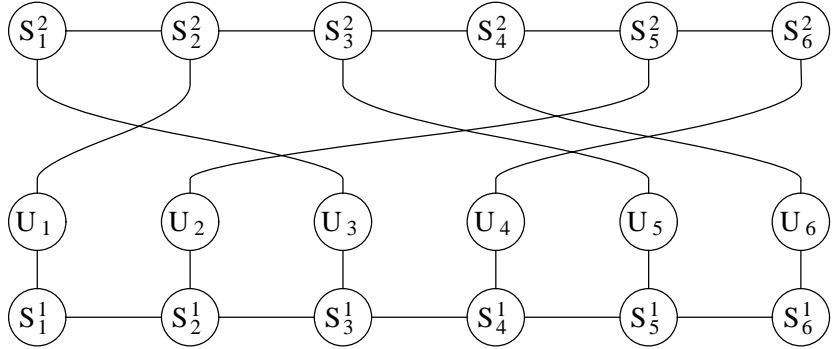}
\end{center}
\caption{The cyclic graph structure underlying the turbo code
of Figure \ref{turbo1}}
\label{turbo2}
\end{figure}

\begin{figure}
\begin{center}
\leavevmode
\includegraphics{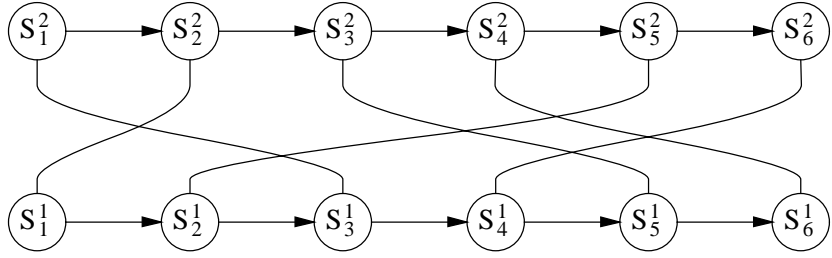}
\end{center}
\caption{The underlying turbo graph for Figure \ref{turbo2}.}
\label{turbo3}
\end{figure}

In Figure \ref{turbo1} the underlying cycle structure is not
affected by the $\mathbf{X}$ and $\mathbf{Y}$ nodes, i.e., they
do not play any role in the counting of cycles in the graph.
For simplicity they can be removed from consideration, resulting
in the simpler graph structure of Figure \ref{turbo2}. Furthermore,
we will drop the directionality of the edges in Figure \ref{turbo2}
and in the rest of the paper, since the definition of a cycle in
an ADG is not a function of the directionality of the edges
on the cycle.

To simplify our analysis further, we initially ignore the nodes
$\mathbf{U_1}$, $\mathbf{U_2}$, $\ldots$,
to arrive at a  {\em turbo graph}  in Figure \ref{turbo3} (we
will later reintroduce the $\mathbf{U}$ nodes).
Formally, a turbo graph is defined as follows:
\begin{enumerate}
\item There are two parallel chains, each having $n$ nodes.
      (For real turbo codes, $n$ can be very large, e.g. $n=64,000$.)
\item Each node is connected to one (and only one) node on the other
    chain and these one-to-one connections are chosen randomly,
        e.g., by a random permutation
        of the sequence $\{1,2,\ldots,n\}$.
 (In Section \ref{sec:srandom} we will look at
        another kind of permutation, the ``S-random  permutation.")
\item A turbo graph as defined above is an {\em undirected} graph.
      But to differentiate between edges on the chains
      and edges connecting nodes on different chains,
      we label the former as being {\em directed} (from left to right),
      and the latter {\em undirected}. (Note: this has nothing
      to do with directed edges in the original ADG model,
      it is just a notational convenience.)
      So an internal node has exactly
      three edges connected to it: one directed edge going out of it,
      one directed edge going into it,
      and one undirected edge connecting it to a node on the other chain.
      A boundary node also has one undirected edge,
      but only one directed edge.
\end{enumerate}

Given a turbo graph, and a randomly chosen node in the graph,
we are interested in:
\begin{enumerate}
 \item counting the number of simple cycles
 of length $k$
which pass through this node
(where a simple cycle is defined as a cycle without repeated nodes), and
 \item finding the probability that this node is not on a simple cycle of
length $k$ or less, for $k=4,5,\ldots$ (clearly the shortest possible cycle in a
turbo graph is 4).
\end{enumerate}

\begin{figure}
    \centering
    \subfigure{
      \label{turbo4}
      \begin{minipage}[b]{.3\textwidth}
          \centering
\includegraphics{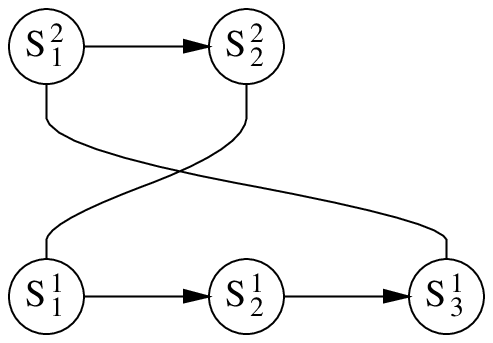}
      \end{minipage}%
     }
    \hspace{0.05in}
    \subfigure{
      \label{turbo5}
      \begin{minipage}[b]{.3\textwidth}
          \centering
\includegraphics{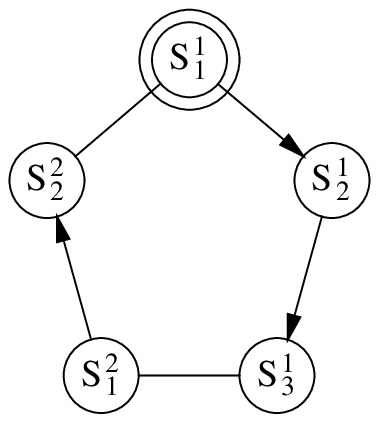}
      \end{minipage}%
     }
    \hspace{0.05in}
    \subfigure{
      \label{turbo6}
      \begin{minipage}[b]{.3\textwidth}
          \centering
\includegraphics{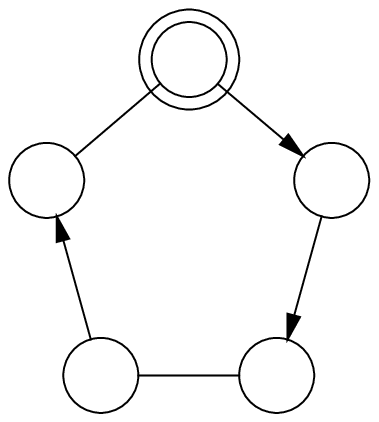}
      \end{minipage}%
     }
    \caption{A simple cycle in Figure \ref{turbo3} and the corresponding
     picture (a) The simple cycle, (b) The same simple cycle, untangled,
and (c) The corresponding picture.}
\end{figure}

\subsection{Embedding ``Pictures"}

To assist our counting of cycles, we introduce
the notion of a {``picture."}
First let us look at Figure \ref{turbo4},
which is a single simple cycle taken from Figure \ref{turbo3}.
When we untangle Figure \ref{turbo4},
we get Figure \ref{turbo5}.
If we omit the node labels, we have Figure \ref{turbo6}
which we call a {\em picture}.

Formally, a picture is defined as follows:
\begin{enumerate}
 \item It is a simple cycle with a single distinguished vertex
       (the circled one in the figure).
 \item It consists of both directed edges and undirected edges.
 \item The number of undirected edges $m$ is even and $m>0$.
 \item No two undirected edges are adjacent.
 \item Adjacent directed edges have the same direction.
\end{enumerate}

We will use pictures as a convenient notation for counting
simple cycles in turbo graphs.
For example, using Figure \ref{turbo6} as a template,
we start from node $\mathbf{S_1^1}$ in Figure \ref{turbo3}.
The first edge in the picture is a directed
forward edge, so we go from $\mathbf{S_1^1}$ along the forward edge
which leads us to $\mathbf{S_2^1}$. The second edge in the picture
is also a directed forward edge, which leads us from $\mathbf{S_2^1}$
to $\mathbf{S_3^1}$.  The next edge is an undirected edge,
so we go from $\mathbf{S_3^1}$ to $\mathbf{S_1^2}$ on the other chain. In the
same way, we go from $\mathbf{S_1^2}$ to $\mathbf{S_2^2}$, then to $\mathbf{S_1^1}$,
which is our starting point. As the path we just traversed
starts from $\mathbf{S_1^1}$ and ends at $\mathbf{S_1^1}$, and there are no
repeated nodes in the path, we conclude that we have found
a simple cycle (of length 5) which is exactly what we
have in Figure \ref{turbo4}.

We can easily enumerate all the
different pictures of length $4,5,...,2n$, and use them
as templates to find all the simple cycles
at a node in a turbo graph. This approach is complete
because any simple cycle in a graph has a corresponding
picture. (To be exact, it has two pictures because
we can traverse it in both directions.)

The process of finding a simple cycle using a picture
as a template can also be thought of
as  {\em embedding}  a picture at a
node in a turbo graph. This embedding may succeed, as in our
example above, or it may fail, e.g., we come to a
previously-visited node other than the starting node,
or we are told to go forward at the end of a chain, etc.
%
%
Using pictures, the problem of counting the number
of simple cycles of length $k$
can be formulated this way:
\begin{itemize}
 \item Count the
number of different pictures of length $k$,
 \item For each distinct picture, calculate the probability of
       embedding it in a turbo graph at a randomly chosen node.
\end{itemize}

\section{Counting Pictures}
\label{sec:count}

We wish to determine the number of different pictures of
length $k$ with $m$ undirected edges.
First, let us define two functions:
\begin{description}
 \item [ $C(a,b)$ ] = the number of ways of picking $b$ disjoint
    edges (i.e., no two edges are adjacent to each other)
    from a cycle of length $a$, with a distinguished vertex and a
    distinguished clockwise direction.
 \item [ $P(a,b)$ ] = the number of ways of picking $b$ independent
    edges from a path of length $a$, with a distinguished endpoint.
\end{description}

These two functions can be evaluated by the following
recursive equations:
\begin{eqnarray}
C(a,b) & = & P(a-1,b) + P(a-3,b-1) \\
P(a,b) & = & P(a-1,b) + P(a-2,b-1)
\end{eqnarray}
and the solutions are
\begin{eqnarray}
P(a,b) & = & \left( \begin{array}{c}
                    a-b+1 \\
                    b
                    \end{array}
             \right) \\
C(a,b) & = & \left( \begin{array}{c}
                  a-b-1 \\
                  b-1
                  \end{array}
           \right)
           +
           \left( \begin{array}{c}
                  a-b \\
                  b
                  \end{array}
           \right)
\end{eqnarray}

Thus, the number of different pictures of length $k$ and with
$m$  undirected
edges,  $0 < m \leq \frac{n}{2}$ (and $m$ is even),
is given by
\begin{eqnarray}
N(k,m) & = &
2^m \left( \left( \begin{array}{c}
                  k-m-1 \\
                  m-1
                  \end{array}
           \right)
           +
           \left( \begin{array}{c}
                  k-m \\
                  m
                  \end{array}
           \right)
     \right) / 2 \nonumber \\
     & = & 2^{m-1} \frac{k}{k-m}
           \left( \begin{array}{c}
                  k-m \\
                  m
                  \end{array}
           \right)
     \label{npic}
\end{eqnarray}

where $2^m$ is the number of different ways to give directions to
the directed edges. The division by two occurs
because  the direction of the picture is irrelevant.
Because of the $m$ undirected edges,
there are $m$ segments of directed edges, with one or more edges
in a segment; the edges within a segment must have a common direction
(property 4 of a picture).

\section{The Probability of Embedding a Picture}
\label{sec:prob}

In this section we derive the
probability $P_n(k,m)$ of embedding a picture of length $k$
and with $m$ undirected edges at a node in a turbo graph
with chain length $n$.

\subsection{When $k=2m$}

Let us first consider a simple picture where the directed edges and
undirected edges alternate (so $k=2m$) and all the directed edges
point in the same (forward) direction.

Let us label the nodes of the picture
as $X_1$,$X_2$,$Y_1$,$Y_2$,
$X_3$,$X_4$,$Y_3$,$Y_4$,$\ldots$,\\
$X_{m-1}$,$X_m$,$Y_{m-1}$,$Y_{m}$.
We want to see if this picture can
be successfully embedded, i.e. if the above nodes
are a simple cycle.
Let us call the chain on which $X_1$ resides {\em side 1},
and the opposite chain {\em side 2}.
The probability of successfully embedding the picture
at $X_1$ is the product of
the probabilities of successfully following each edge of
the picture, namely,
\begin{itemize}
 \item $X_1 \rightarrow X_2$. This will fail if $X_1$ is
  the right-most node on side 1. So $p=1-\frac{1}{n}$.

 \item $X_2 \rightarrow Y_1$. Here $p=1$.

 \item $Y_1 \rightarrow Y_2$. This will fail if $Y_1$ is
  the right-most node on side 2. So $p=1-\frac{1}{n}$.

 \item $Y_2 \rightarrow X_3$.
  $X_3$ is the ``cross-chain" neighbor of $Y_2$. As there is already
  a connection between $X_2$ and $Y_1$, $X_3$ cannot possibly
  coincide with $X_2$; but it may coincide with $X_1$ and
  make the embedding fail. This gives us $p=1-\frac{1}{n-1}$.

  More generally, if there are $2s$ visited nodes on side 1,
  then $s$ of them already have their connections to side 2.
  So from a node on side 2, there are only $n-s$ nodes
  on side 1 to go to, $s$ of which are visited nodes.
  So $p=1-\frac{s}{n-s}$.

 \item $X_3 \rightarrow X_4$.
  Here we have two previously visited nodes ($X_1$,$X_2$).
  When there are $2s$ previously-visited nodes,
  the unvisited nodes are partitioned into up to $s$ segments,
  and after we come from side 2 to side 1, if we fall
  on the right-most node of one of the segments, the embedding will fail:
  either we go off the chain, or we go to a previously-visited node.
  In this way, we have $1-\frac{s+1}{n-2s} \leq p \leq 1-\frac{1}{n-2s}$.

 \item $\cdots$

 \item $Y_{m-1} \rightarrow Y_m$.

 \item $Y_m \rightarrow X_1$. $p=\frac{1}{n-\frac{m}{2}}$.
  This final step ( $Y_m \rightarrow X_1$) completes the cycle.

\end{itemize}

Multiplying these terms, we arrive at
\begin{eqnarray}
& &
\frac{1}{n-\frac{m}{2}}
  \prod_{s=0}^{s=\frac{m}{2}}
    \left[  \left(1-\frac{s}{n-s}\right)
      \left(1-\frac{s+1}{n-2s}\right)
    \right]^2 \nonumber \\
&\leq&  P_n(k,m) \\
& \leq &
\frac{1}{n-\frac{m}{2}}
  \prod_{s=0}^{s=\frac{m}{2}}
    \left[  \left(1-\frac{s}{n-s}\right)
      \left(1-\frac{1}{n-2s}\right)
    \right]^2 \nonumber
\end{eqnarray}

For large $n$  and small $m$,
the ratio between the upper bound and the lower bound is close to 1.
For example, when $n=64,000$ and $m=10$ the ratio is $1.0005$.

\subsection{The general case}

The above analysis can be extended easily to the general case
where:
\begin{itemize}
  \item The directed edges in the picture are
not constrained to be unidirectional.
  \item $k \geq 2m$. (Because the $m$ undirected
     edges cannot be adjacent to each other, the
     total number of edges $k$ must be $\geq 2m$.)
\end{itemize}

When $k=2m$, no two directed edges are adjacent. Equivalently,
there are $m$ segments of directed edges, and in each segment,
there is only one edge. When $k > 2m$, we still have $m$ segments
of directed edges, but there is more than one edge in a segment.
Suppose for $1 \leq i \leq m$, the $i$th segment of side 1
has $a_i$ edges, and the $i$th segment of side 2 has $b_i$ edges.
$P_{n}(k,m)$ is given by:

\begin{eqnarray}
& &
\frac{1}{n-\frac{m}{2}}
  \prod_{s=0}^{s=\frac{m}{2}}
    \left[  \left(1-\frac{\sum_{i=1}^{s}a_i }{n-s}\right)
      \left(1-\frac{s+1}{n-\sum_{i=1}^{s}(a_i+1)}\right)
            \left(1-\frac{\sum_{i=1}^{s}b_i }{n-s}\right)
      \left(1-\frac{s+1}{n-\sum_{i=1}^{s}(b_i+1)}\right)
    \right] \nonumber
\\
& \leq & P_n(k,m)
\label{bound1}
\\
& \leq &
\frac{1}{n-\frac{m}{2}}
  \prod_{s=0}^{s=\frac{m}{2}}
    \left[  \left(1-\frac{\sum_{i=1}^{s}a_i }{n-s}\right)
      \left(1-\frac{1}{n-\sum_{i=1}^{s}(a_i+1)}\right)
            \left(1-\frac{\sum_{i=1}^{s}b_i }{n-s}\right)
      \left(1-\frac{1}{n-\sum_{i=1}^{s}(b_i+1)}\right)
    \right] \nonumber
\end{eqnarray}

From
\[
\sum_{i=1}^{m} a_i + \sum_{i=1}^{m} b_i = k-m,
\]

\[
1 \leq a_i,\,b_i \leq 1+(k-2m),
\]

\[
\sum_{i=1}^{s} a_i+\sum_{i=1}^{s} b_i \leq 2s+(k-2m),
\]
and
\[
s \leq \sum_{i=1}^{s} a_i, \, \sum_{i=1}^{s} b_i \leq s+(k-2m)
\]
we can simplify the bounds in Equation \ref{bound1} to
\begin{eqnarray}
& &
\frac{1}{n-\frac{m}{2}}
  \prod_{s=0}^{s=\frac{m}{2}}
    \left[  \left(1-\frac{s+k-2m}{n-s}\right)
      \left(1-\frac{s+1}{n-(2s+k-2m)}\right)
    \right]^2 \nonumber
\\
& \leq  & P_n(k,m)
\label{bounds}
\\
 & \leq &
\frac{1}{n-\frac{m}{2}}
  \prod_{s=0}^{s=\frac{m}{2}}
    \left[  \left(1-\frac{s}{n-s}\right)
      \left(1-\frac{1}{n-2s)}\right)
    \right]^2 \nonumber
\end{eqnarray}
The ratio between the upper bound and the lower bound is still
close to 1. For example, when $n=64,000, k=10, m=4$,
the ratio is $1.0003$. Given that the
bounds are so close in the range of $n, k$, and $m$
of interest, in the remainder of the
paper we will simply approximate $P_n(k,m)$ by the arithmetic
average of the upper and lower bound.

\section{The Probability of No Cycles of Length $k$ or Less}
\label{sec:nocycles}

In Section \ref{sec:count} we derived $N(k,m)$,
the number of different pictures
of length $k$ with $m$ undirected cycles (Equation (\ref{npic})).
In Section \ref{sec:prob} we  estimated $P_n(k,m)$,
the probability of embedding a picture (with length $k$
and $m$ undirected edges) at a node in a turbo graph
with chain length $n$ (Equation (\ref{bounds})).
With these two results, we can now determine the
probability of no cycle of length $k$ or
less at a randomly chosen node in a turbo graph of length $n$.

Let $P(\overline{{\cal L}_k})$ be the probability
that there are no cycles of length $k$ at a randomly chosen
node in a turbo graph.
Thus,
\begin{eqnarray}
 P(no\ cycle\ of\ length\ \leq k)
 & = & P(\overline{{\cal L}}_k, \overline{{\cal L}}_{k-1},\ldots, \overline{{\cal L}}_4) \nonumber \\
 & = & \prod_{i=4}^{k} P(\overline{{\cal L}}_i \mid \overline{{\cal L}}_{i-1}, \ldots, \overline{{\cal L}}_4) \nonumber \\
 & \approx & \prod_{i=4}^{k} P(\overline{{\cal L}}_i)
 \label{independence1}
\end{eqnarray}
In this independence approximation we are assuming that
at any particular node
the event ``there are no cycles of length $k$" is independent
of the event ``there are no cycles of length $k-1$ or
lower." This is not strictly true since (for example) the non-existence
of a cycle of length $k-1$ can make
certain cycles of length $k$ impossible (e.g., consider the case $k=5$).
However, these cases appear to be relatively rare, leading us
to believe that the independence assumption is relatively
accurate to first-order.


Now we estimate $ P(\overline{{\cal L}}_k)$, the probability
of no cycle of length $k$ at a randomly chosen node.
Denote the individual pictures of length $k$ as
$pic_1$,$pic_2$,$\ldots$, and let $\overline{pic}_i$ mean that
the $i$th picture fails to be embedded.

\begin{eqnarray}
  P(\overline{{\cal L}}_k)
 & = & P( \overline{pic}_1, \overline{pic}_2,  \ldots) \nonumber \\
 & = & \prod_i P( \overline{pic}_i \mid \overline{pic}_{i-1}, \ldots, \overline{pic}_{1}) \nonumber \\
 & \approx & \prod_{i} P(\overline{pic}_i) \nonumber \\
 & = & \prod_{m>0,m \ even}^{m \leq \frac{k}{2}}
     \left(1-P_n(k,m)\right)^{N(k,m)}
\label{maineqn}
\end{eqnarray}
Here we make a second independence assumption which again may
be violated in practice.   The non-existence of embedding
of certain pictures (the event
being conditioned on) will influence the probability
of existence of embedding of other pictures. However, we
conjecture
that this dependence is rather weak and that the independence
assumption is again a  good first-order approximation.


\section{Numerical and Simulation Results}
\label{sec:simulation}

\begin{figure}
\centering
\subfigure[]{
\begin{tabular}{||r|r|r|r||r||}
\hline
 k & $P_{simulation}$ & $P_{theoretical}$ & Difference & $\hat{\sigma}_{P}$ \\
\hline
 4 & 0.999950 & 0.999938 &  0.000012  & 0.000056  \\
 5 & 0.999750 & 0.999781 & -0.000031  & 0.000105  \\
 6 & 0.999450 & 0.999500 & -0.000050  & 0.000158  \\
 7 & 0.999100 & 0.999063 &  0.000037  & 0.000216  \\
 8 & 0.998350 & 0.998189 &  0.000161  & 0.000301  \\
 9 & 0.996650 & 0.996227 &  0.000423  & 0.000434  \\
10 & 0.992400 & 0.992034 &  0.000366  & 0.000629  \\
11 & 0.983750 & 0.983886 & -0.000136  & 0.000890  \\
12 & 0.968400 & 0.968456 & -0.000056  & 0.001236  \\
13 & 0.938850 & 0.938643 &  0.000207  & 0.001697  \\
14 & 0.881800 & 0.880781 &  0.001019  & 0.002291  \\
15 & 0.775350 & 0.774188 &  0.001162  & 0.002957  \\
16 & 0.600550 & 0.598375 &  0.002175  & 0.003466  \\
17 & 0.358850 & 0.358868 & -0.000018  & 0.003392  \\
18 & 0.125850 & 0.129488 & -0.003638  & 0.002374  \\
19 & 0.015500 & 0.016782 & -0.001282  & 0.000908  \\
20 & 0.000150 & 0.000279 & -0.000129  & 0.000118  \\
\hline
\end{tabular}
}
\subfigure[]{
\includegraphics[width=4.5in,height=3in]{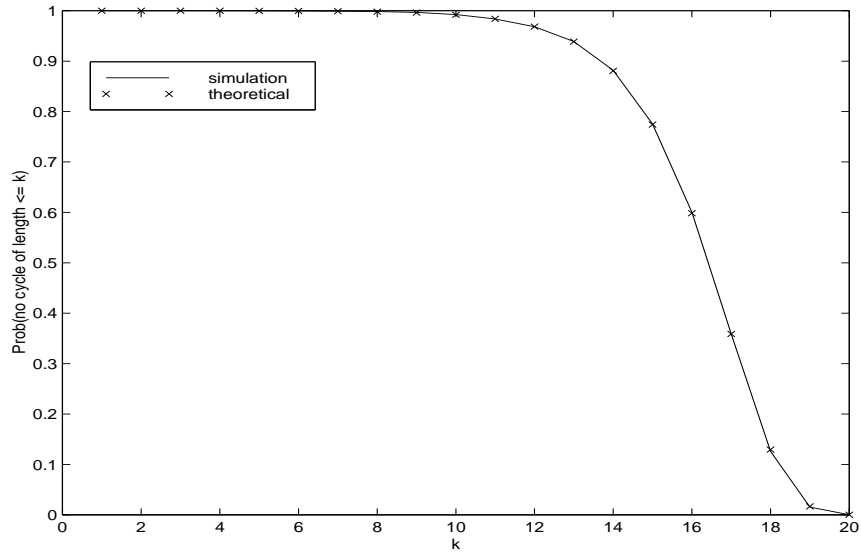}
}
\caption{Theoretical vs. simulation estimates of
the probability of no cycles of length $k$ or less,
as a function of $k$.
Turbo graph chain length $n=64,000$.}
\label{appxtable}
\end{figure}

\subsection{Cycle Length Distributions in Turbo Graphs}

We ran a series of simulations where
200 different turbo graphs (i.e., each graph
has a different random permuter) of length $n=64000$
are randomly generated. For each graph,
we counted the simple cycles of length $k=4,5,\ldots,20$,
at 100 randomly chosen nodes.
In total, the cycle counts at $20000$ nodes are collected
to generate an empirical estimate of
the true $P(no~cycle~of~length~\leq~k)$.
The theoretical estimates are derived by
using the independence assumptions of
Equations (\ref{independence1}) and (\ref{maineqn}).
 $P_n(k,m)$ is calculated  as the arithmetic
average of  the two bounds
in Equation (\ref{bounds}).

The simulation results, together with the theoretical estimates
are shown in Figure \ref{appxtable}.  The difference in
error is never greater than about 0.005 in probability.
Note that neither the sample-based estimates nor
the theoretical estimates are exact. Thus, differences between
the two could be due to either sampling variation or error introduced
by the independence assumptions in the estimation. The fact
that the difference in errors is non-systematic (i.e., contains
both positive and negative errors) suggests that both methods
of estimation are fairly accurate. For comparison, in the last column
of the table we provide the estimated standard deviation
 $\hat{\sigma}_{P}=\sqrt{\hat{P}(1-\hat{P})/N}$, where
$\hat{P}$ is the simulation estimate.
We can see that the differences between $P_{simulation}$ and $P_{theoretical}$
are within $\pm \hat{\sigma}_{P}$ of $P_{theoretical}$ except for
the last three rows where $P_{theoretical}$ is quite small. For
large $k$ we can expect that the simulation
estimate of $\hat{P}$ will be less accurate since we
are estimating relatively rare events. Thus, since
our estimate of
 $\hat{\sigma}_{P}$ is a function of  $\hat{P}$, for larger
$k$ values any  differences between theory and simulation
could be due entirely to sampling error.


\begin{figure}
\begin{center}
\leavevmode
\includegraphics[width=4.5in]{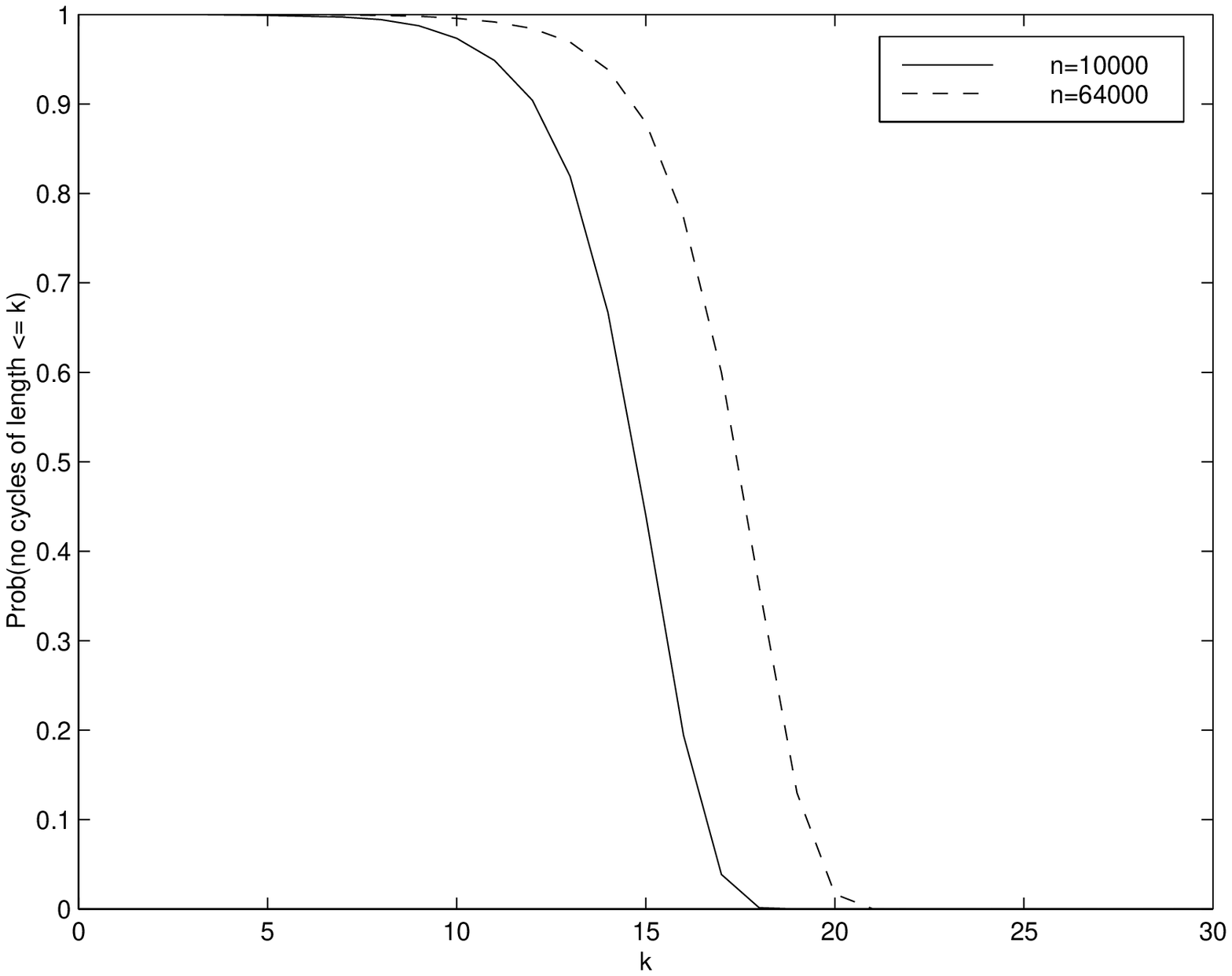}
\end{center}
\caption{ Approximate probability of no cycles of length $k$ or less,
as a function of $k$.}
\label{numericresult}
\end{figure}

Figure \ref{numericresult} shows a plot of the estimated probability
that there are no cycles of length $k$ or less
at a randomly chosen node.
There appears to be a ``soft threshold
effect" in the sense that beyond a certain value of $k$,
it rapidly becomes much more likely that there are cycles
of length $k$ or less at a randomly chosen node.
The location
of this threshold increases as $n$ increases (i.e., as the
length of the chain gets longer).

\begin{figure}
\begin{center}
\leavevmode
\includegraphics[width=4.5in]{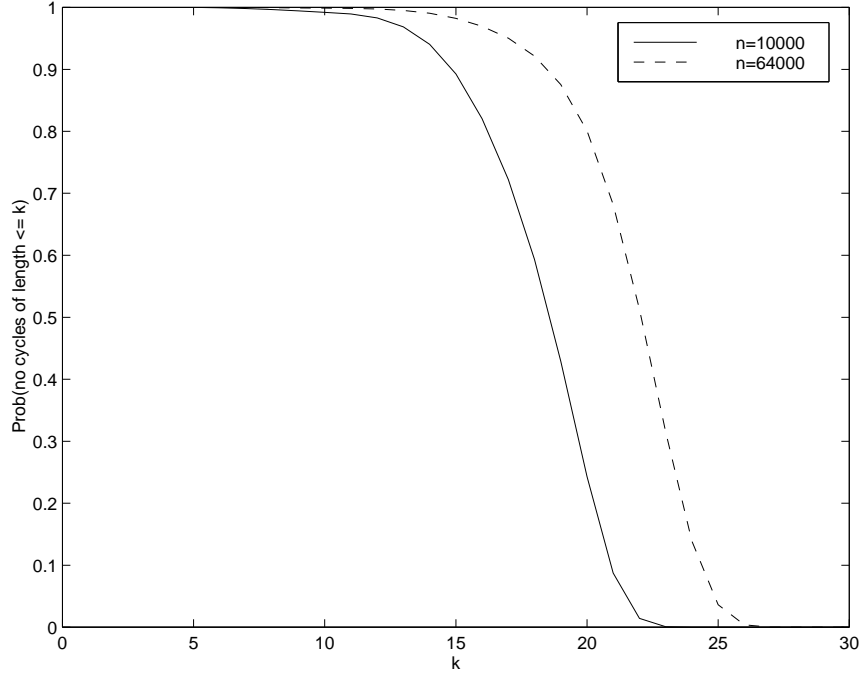}
\end{center}
\caption{Probability of no cycles of length $k$ or less,
including the $\mathbf{U}$ nodes
(Figures \ref{turbo2})
in the ADG for turbo decoding,
as a function of $k$.}
\label{numericu}
\end{figure}

\subsection{Large-Sample Closed-Form Approximations}


When $n$ is sufficiently large, (i.e., $n \gg k$),
the probability of embedding a picture (Equation (\ref{bounds}))
can simply written as
\begin{equation}
P_n \approx \frac{1}{n}
\end{equation}
In this case, we do not differentiate between pictures
with different numbers of undirected edges

The total number of pictures of length $k$ is
\begin{eqnarray}
N_k & = & \sum_{m>0,m\ even}^{m \leq \frac{k}{2}} N(k,m) \nonumber \\
&=&  \sum_{m>0,m\ even}^{m \leq \frac{k}{2}} 2^{m-1}\frac{k}{k-m}
 \left( \begin{array}{c}
                  k-m \\
                  m
                  \end{array}
           \right) \nonumber \\
 &\approx & 2^{k-2}
\end{eqnarray}

The log probability of no cycle of length $k$ is then
\begin{equation}
\log P(\overline{{\cal L}}_{k})
 \approx   \log (1-P_n)^{N_k}
\approx 2^{k-2} \log (1-\frac{1}{n})
\approx -\frac{1}{n} 2^{k-2},
\end{equation}
from which one has
\begin{eqnarray}
 \log P(no\ cycle\ of\ length\ \leq k)
 & \approx & \log
\left(\prod_{i=4}^{k} P(\overline{{\cal L}}_{i}) \right) \nonumber \\
 &  \approx & \sum_{i=4}^{k} \left( -\frac{1}{n} 2^{i-2} \right) \nonumber \\
 &  = & -\frac{2^{k-1}-4}{n}
\end{eqnarray}

Thus, the probability of no cycle of length $k$ or less
is approximately $e^{-\frac{2^{k-1}-4}{n}}, k \ge 4$.
This probability equals 0.5
at $k_{0.5}= \log_2 (n\log 2 +4) + 1$, which provides an
indication of how the curve will shift to the right
as $n$ increases. Roughly speaking, to double $k_{0.5}$,
one would  have to square the block-length
of the code from $n$ to $n^2$.

\subsection{Including the $\mathbf{U}$ Nodes}
Up to this point we have been counting cycles in
the turbo graph (Figure \ref{turbo3})
where we ignore the  information nodes,
$U_i$.
The results can readily
be extended to include these $\mathbf{U}$ nodes
by counting each undirected edge (that connects nodes from different chains)
as two edges.


Let $m'=\frac{m}{2},\ k'=k-\frac{m}{2}$ be the number
of undirected edges and the cycle length, respectively,
when we ignore the $\mathbf{U}$ nodes.
From $m'>0,\ m'  \ even,\ m' \leq \frac{k'}{2}$,
we have $m>0,\ m \ divisible\ by\ 4,\ m \leq \frac{2k}{3}$.

Substituting these into Equation \ref{maineqn}, we have
\begin{eqnarray}
 & &   \prod_{m'>0,m' \ even}^{m' \leq \frac{k'}{2}}
     \left(1-P_{n}(k',m')\right)^{N(k',m')} \nonumber \\
 & = &
\prod_{m>0,m \ divisible\ by\ 4}^{m \leq \frac{2k}{3}}
     \left(1-P_{n}
\left(k-\frac{m}{2},\frac{m}{2}\right)\right)^{N(k-\frac{m}{2},\frac{m}{2})}
\label{includeu}
\end{eqnarray}

Using Equation \ref{includeu}, we plot in Figure \ref{numericu}
the estimated probability of no cycles of length $k$ or
less in the graph for turbo decoding which
includes the $\mathbf{U}$ nodes (Figure \ref{turbo2}). Not
surprisingly, the effect is to ``shift" the graph to the
right, i.e., adding $\mathbf{U}$ nodes has the effect of
lengthening the typical cycle.

For the purposes of investigating the properties
of the message-passing
algorithm, the relevant nodes on a cycle may well
be those which are directly connected to a $\mathbf{Y}$
node (for example, the $\mathbf{U}$ nodes in a systematic
code and any $\mathbf{S}$ nodes which are producing
a transmitted codeword). The rationale for including these
particular nodes (and not including nodes which are
not connected to a $\mathbf{Y}$ node) is that these
are the only ``information nodes"
in the graph  that  in effect can transmit messages
that potentially  lead to multiple-counting. It
is possible that it is only
the number of these nodes on a cycle which is relevant
to message-passing algorithms.  Thus,
for a particular code structure, the relevant nodes to count
in a cycle could be redefined to be
only those which have an associated $\mathbf{Y}$. The general
framework we have presented here can easily be modified to
allow for such counting.

Note  also that various extensions of turbo codes are
also amenable to this form of analysis. For example, for
the case of a turbo code with more than two constituent
encoders, one can generalize the notion of a picture and
count accordingly.

\section{The ``S-random" permutation}
\label{sec:srandom}

In our construction of the turbo graph (Figure \ref{turbo3})
we use a random permutation,
i.e. the one-to-one connections of nodes from the two chains
are chosen randomly by a random permutation.
In this section we look at the
``S-random" permutation [10], a particular semi-random
construction.


%
%
%
%

\begin{figure}
\begin{center}
\leavevmode
\includegraphics{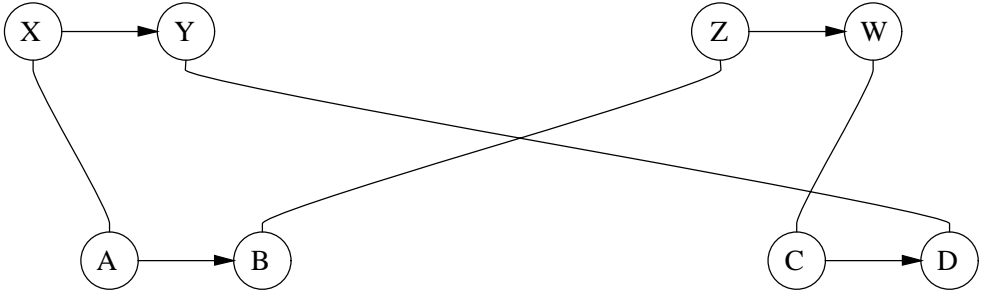}
\end{center}
\caption{A cycle of length 8}
\label{srand8}
\end{figure}

Formally, the S-random permutation
is a random permutation function $f(\cdot)$ on the sequence
$1,2,\ldots,n$ such that
\begin{equation}
\forall i,j: \mid i -j \mid \leq S
\,\, \Longrightarrow \,\,
\mid f(i)-f(j) \mid \geq S
\end{equation}
The S-random permutation stipulates that if two
nodes on a chain are within a distance $S$ of each other,
their counterparts on the other chain cannot be within
a distance $S$ of each other.
This restriction will eliminate some of the cycles occurring
in a turbo graph with a purely random permutation.
For example, there cannot be any cycles in the graph of
length $k=$4, 5, 6 or 7.
Thus, the S-random construction disallows
cycles of length $k$ for $k < 8$. However, from
Section \ref{sec:simulation} we know that these short cycles ($k < 8$)
occur relatively rarely in realistic turbo codes.
In Figure \ref{srand8}, we show a cycle of length $k=8$.
As long as the distances of $\mathbf{\mid YZ \mid}$
and $\mathbf{\mid BC \mid}$ are large enough ($>S$),
cycles of lengths $k \ge  8$ are possible for any $S$.

\begin{table}
\begin{center}
\begin{tabular}{||r||c||r|r|r|r||} \hline
  \multicolumn{6}{|c|}{Prob(no cycle of length $k$ or less)} \\
  \multicolumn{6}{|c|}{for turbo graph ($n=64000$)} \\ \hline
 & Random &  \multicolumn{4}{|c||}{S-random permutation} \\ \cline{3-6}
k &  permutation  & $S=10$ & $S=20$ & $S=50$ & $S=100$ \\ \hline
 4 & 1.0000 & 1.0000 & 1.0000 & 1.0000 & 1.0000 \\
 5 & 0.9998 & 1.0000 & 1.0000 & 1.0000 & 1.0000 \\
 6 & 0.9995 & 1.0000 & 1.0000 & 1.0000 & 1.0000 \\
 7 & 0.9991 & 1.0000 & 1.0000 & 1.0000 & 1.0000 \\
 8 & 0.9984 & 0.9996 & 0.9998 & 0.9998 & 0.9998 \\
 9 & 0.9967 & 0.9983 & 0.9987 & 0.9987 & 0.9984 \\
10 & 0.9924 & 0.9949 & 0.9945 & 0.9956 & 0.9950 \\
11 & 0.9838 & 0.9890 & 0.9891 & 0.9877 & 0.9887 \\
12 & 0.9684 & 0.9739 & 0.9765 & 0.9736 & 0.9748 \\
13 & 0.9389 & 0.9460 & 0.9503 & 0.9449 & 0.9478 \\
14 & 0.8818 & 0.8877 & 0.8920 & 0.8904 & 0.8913 \\
15 & 0.7754 & 0.7804 & 0.7847 & 0.7858 & 0.7833 \\
16 & 0.6006 & 0.6114 & 0.6014 & 0.6121 & 0.6006 \\
17 & 0.3589 & 0.3671 & 0.3629 & 0.3731 & 0.3647 \\
18 & 0.1259 & 0.1315 & 0.1289 & 0.1360 & 0.1330 \\
19 & 0.0155 & 0.0146 & 0.0164 & 0.0184 & 0.0183 \\
20 & 0.0002 & 0.0004 & 0.0003 & 0.0004 & 0.0008 \\
\hline
\end{tabular}
\caption{Simulation-based estimates of the probability of no cycle
of length $k$ or less, comparing the standard random
construction with the S-random construction.}
\label{tab:srandom}
\end{center}
\end{table}

We simulated S-random graphs and counted
cycles in the same manner
as described in Section \ref{sec:simulation}, except that
the random permutation was now carried out in
the S-random fashion as described in [10]. The results in
 Table \ref{tab:srandom} show that changing the value
of $S$ does not appear to significantly change the nature of the
cycle-distribution. The S-random distributions of course have zero
probability for $k< 8$, but for $k \ge 8$ the results from both
types of permutation appear qualitatively similar, with a
 small systematic increase in the probability of a node
not having a cycle of length $k$ for the S-random case (relative
to the purely random permutation). As the cycle-length $k$
increases, the difference between the S-random and random
distributions narrow. For relatively short cycles
with values of $k$ between 8 and 12 (say) the difference
is relatively substantial if one considers the 
the probability of {\it having} a cycle of length less
than or equal to $k$. For example, for $k=10 $ and $S=100$,
the S-random probability is 0.0050 while the probability
for the random permuter is 0.0076 (see Table \ref{tab:srandom}).

In [11, 12] it was shown (empirically) that the S-random
construction does not have an ``error floor" of the form
associated with a random graph, i.e., the probability of bit error
steadily decreases with increasing SNR for the S-random
construction. The improvement in bit error rate is attributed to
  the improved
weight distribution properties of the
code resulting from the S-random construction. From
a cycle-length viewpoint the S-random construction essentially
only differs slightly from the random construction (e.g., by
eliminating the relatively rare cycles of length $k=4, 5, 6$ and
$7$). Note, however, that because two graphs have very similar
cycle length distributions does not necessarily imply that they
will have similar coding performance. It is possible
that the elimination of the very short cycles combined with the
small systematic increase in the probability of not having a cycle
of length $k$ or less ($k \ge 8$), may be a contributing factor in the
observed improvement in bit error rate, i.e., that even a small
systematic reduction in the number of short  cycles in the graph may
translate into  the empirically-observed
 improvement in coding performance.

\section{Low-Density Parity Check Codes}
\label{sec:ldpc} LDPC codes are another
class of codes exhibiting characteristics and performance similar
to turbo codes [13, 14]. Like turbo codes, the underlying ADG has
loops, rendering exact decoding intractable. Once again, however,
iterative decoding (aka message-passing) works well in practice.
Recent analyses of iterative decoding for LDPC codes have assumed
that there are no short cycles in the LDPC graph structure [15,
16]. Thus, as with turbo codes, it is again of interest to
investigate the distribution of cycle lengths for realistic LDPC
codes.

\begin{figure}
\begin{center}
\leavevmode
\includegraphics{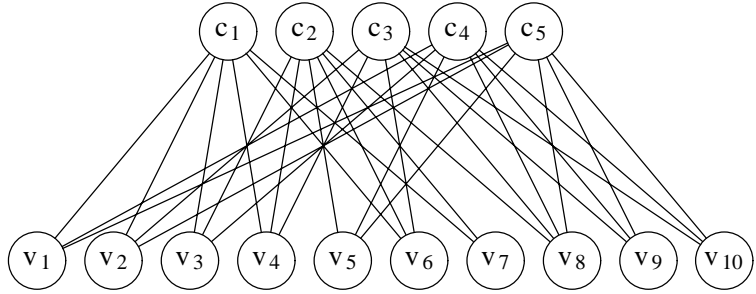}
\end{center}
\caption{Graph structure of Low-Density Parity Check Codes:
$d_v=3,d_c=6,n=10$.}
\label{ldpc}
\end{figure}

The graph structure of regular LDPC codes is shown in Figure \ref{ldpc}
(an {\it LDPC graph}). In this bipartite graph, at the bottom are
$n$ variable nodes $\mathbf{v_1}, \mathbf{v_2}, \ldots,
\mathbf{v_n}$, and at the top are the $w$ check nodes
$\mathbf{c_1}, \mathbf{c_2}, \ldots,\mathbf{c_w}$. For the regular
random LDPC construction each variable node has degree $d_v$,
 each check node has degree $d_c$ (obviously $n d_v = w
d_c$), and the connectivity is generated in a random fashion.

Using our notion of a {\it picture}, we can also analyze
the distribution of cycle lengths in LDPC graphs as we have
done in turbo graphs.  Obviously, here the cycle length must
be even.

We define a picture for an LDPC graph as
follows. Recall that in a turbo
graph, the edges in a picture are labeled as {\it undirected,
forward}, or {\it backward}. For an LDPC graph, we label an edge
in a picture by a number $i$ between 1 and $d_v$
(or between 1 and $d_c$)
to denote that this edge is the $i$-th edge coming from
a node.


First consider the probability of successfully embedding
a picture of length $k=2m$ at a randomly chosen node in an LDPC graph.

\begin{eqnarray}
 P_{embed}(k=2m)
& = & 1 \cdot (1-\frac{1}{d_c}) \cdot (1-\frac{1}{d_v}) \nonumber \\
& &  \cdot \left[ (1-\frac{1}{d_c})(1-\frac{1}{n-1}) \right] \nonumber \\
& &  \cdot \left[ (1-\frac{1}{d_v})(1-\frac{1}{c-1}) \right] \nonumber \\
& &  \cdot \left[ (1-\frac{1}{d_c})(1-\frac{2}{n-1}) \right] \nonumber \\
& &  \cdot \left[ (1-\frac{1}{d_v})(1-\frac{2}{c-1}) \right] \nonumber \\
& & \cdots \nonumber \\
& &  \cdot \left[ (1-\frac{1}{d_c})(1-\frac{m-2}{n-1}) \right] \nonumber \\
& &  \cdot \left[ (1-\frac{1}{d_v})(1-\frac{m-2}{c-1}) \right] \nonumber \\
& &  \cdot \left[ (1-\frac{1}{d_c})\frac{1}{n-1} \right] \nonumber \\
& = &
\frac{1}{n-1}
\left( 1-\frac{1}{d_c} \right) ^{m}
\left( 1-\frac{1}{d_v} \right) ^{m-1}
\prod_{i=0}^{m-2}
\left[
    \left( 1-\frac{i}{n-1} \right)
    \left( 1-\frac{i}{c-1} \right)
\right] \nonumber
\end{eqnarray}

The number of different pictures of length $k=2m$ is
\begin{equation}
 N(k=2m) = \frac{1}{2} d_c^m d_v^m
\end{equation}

Finally, the probability of no cycle
of length $k=2m$ at a randomly chosen node in a LDPC graph is:
\begin{eqnarray}
& & Prob(no\ cycle\ of\ length\ k=2m\ or\ less) \nonumber \\
& \approx & \prod_{i=4,i\ even}^{k} Prob(no\ cycle\ of\ length\ i)
\label{ldpceqn}
\\
& \approx & \prod_{i=4,i\ even}^{k} \left(1 - P_{embed}(i)\right)^{N(i)} \nonumber
\end{eqnarray}
where we make the same two independence assumptions as we did for the
turbo code case.


\begin{figure}
\begin{center}
\leavevmode
\includegraphics[width=4.5in]{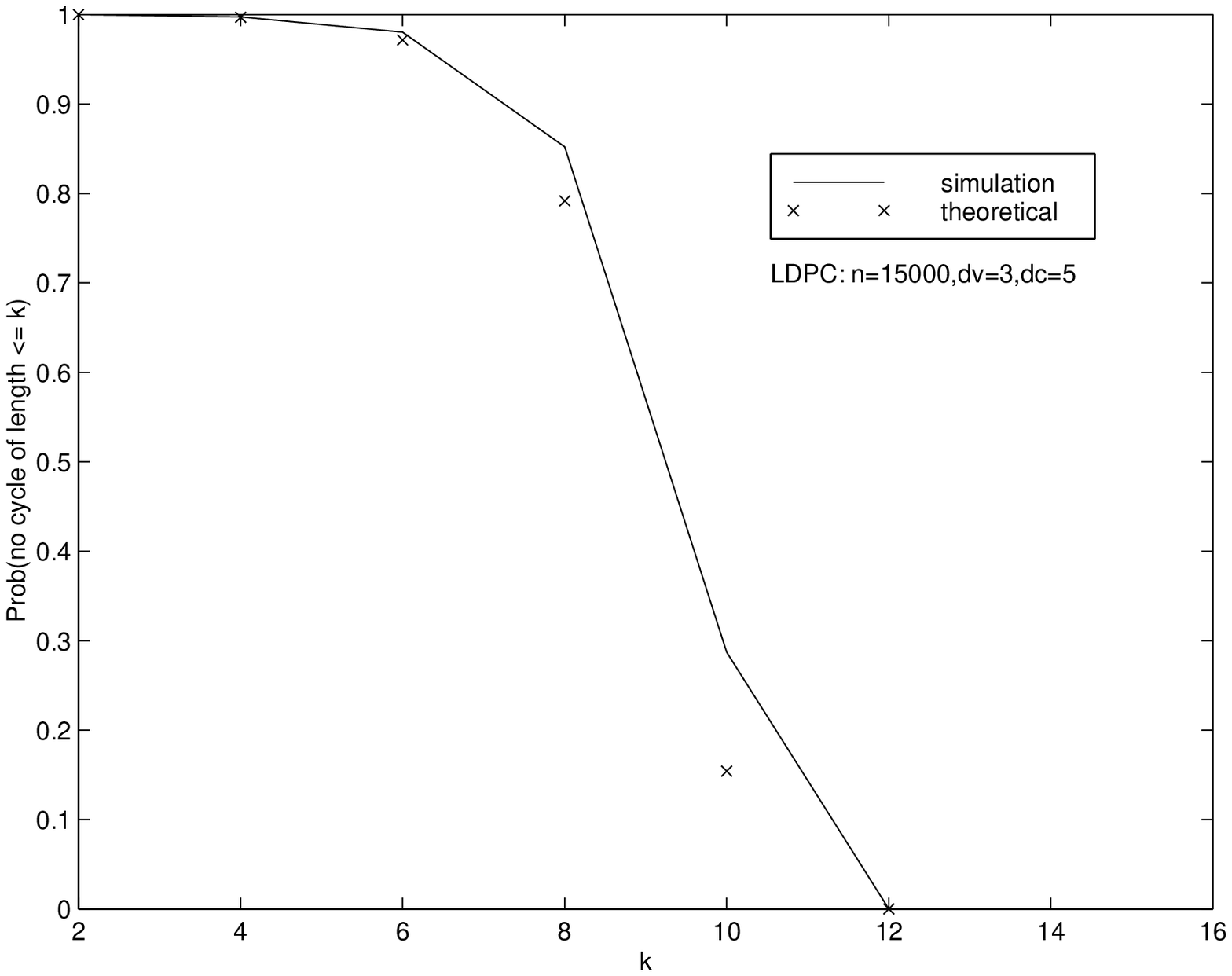}
\end{center}
\caption{The probability of no cycles of length $k$ or less
in an LDPC graph with $n=15000$,
as a function of $k$.}
\label{ldpc15k}
\end{figure}

\begin{figure}
\begin{center}
\leavevmode
\includegraphics[width=4.5in]{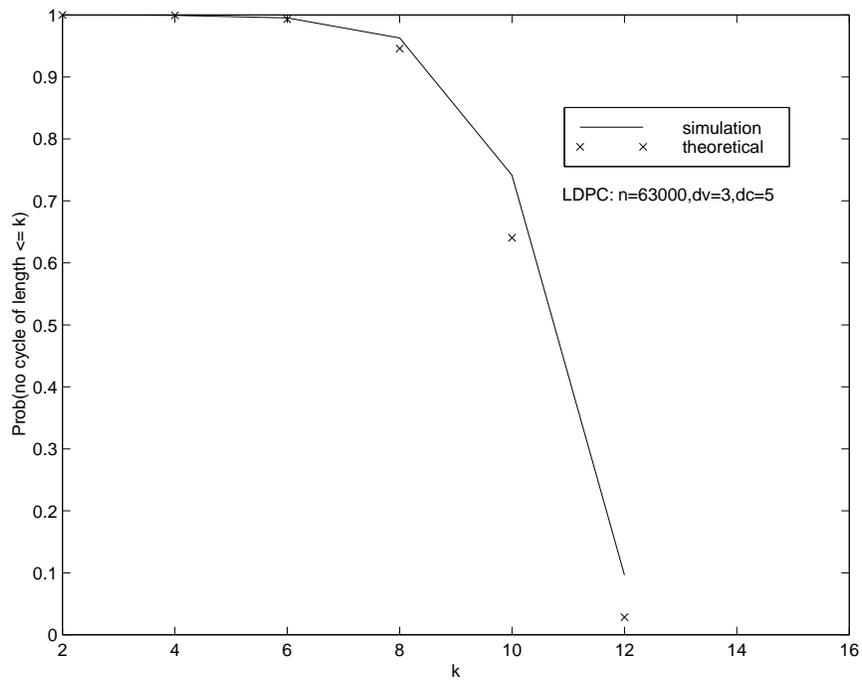}
\end{center}
\caption{The probability of no cycles of length $k$ or less
in an LDPC graph with $n=63000$,
as a function of $k$.}
\label{ldpc63k}
\end{figure}

We ran a number of simulations in which we randomly generated
200 different randomly generated
LDPC graphs and counted the cycles at 100
randomly chosen nodes in each.  We plot in Figures
\ref{ldpc15k} and \ref{ldpc63k} the results of the simulation and the
theoretical estimates from Equation \ref{ldpceqn} for $n=15000$ and 63000.

From the simulation results we see that the LDPC
curve is qualitatively similar in shape to the turbo graph
curves earlier but has been shifted to the left, i.e.,
there is a higher probability of short cycles in an
LDPC graph than in a turbo graph, for the specific parameters
we have looked at here. This is not surprising
since the branching factor in a turbo graph is 3 (each
node is connected to 3 neighbors) while the average branching
factor in an LDPC graph (as analyzed with $d_c=5, d_v = 3$)
is 4.

Existing theoretical analyses of the message-passing
algorithms for LDPC codes rely on the
assumption that none of
the cycles in the underlying graph are short [e.g., 15, 16]. In
contrast, here we explicitly estimate the distribution
on cycle lengths, and find (e.g., Figure 10 and 11)
that there is a ``soft threshold"
effect (as with turbo graphs). For example, for $n=15000, d_v = 3, d_c = 5$,
the simulation results in Figure 10 illustrate that
the probability is about 50\% that a randomly
chosen node participates in a simple cycle of length 9 or
less.

The independence assumptions clearly are not
as accurate in the LDPC case as they were for the
turbo graphs. Recall that we make two separate
independence assumptions in our analysis, namely that
\begin{enumerate}
\item the event that there is no cycle of length $k$
 is independent of the event that there are no cycles of length
$k-1$ or lower, and
\item the event that a particular picture cannot be embedded
at a randomly chosen node is independent of the event
that other pictures cannot be embedded.
\end{enumerate}

We can check the accuracy
of the first independence assumption readily
by simulation.
We ran a number of simulations to count cycles in randomly generated
turbo and LDPC
graphs. From the simulation data, we estimate the marginal probabilities
$P(\overline{{\cal L}}_k)$, and the joint probabilities $P(\overline{{\cal
L}}_k, \overline{{\cal L}}_{k+1})$.  To test the accuracy
of our independence assumption, we compare the product of the
estimated marginal probabilities
with the estimated joint probability.

\begin{table}
\begin{center}
\begin{tabular}{|r|r|r|r|}
\hline
 k & $P(\overline{{\cal L}}_k)P(\overline{{\cal L}}_{k+1})$
 & $P(\overline{{\cal L}}_k, \overline{{\cal L}}_{k+1})$  & Difference \\
\hline
 1 & 1.000000 & 1.000000 & 0.000000 \\
 2 & 1.000000 & 1.000000 & 0.000000 \\
 3 & 0.999950 & 0.999950 & 0.000000 \\
 4 & 0.999750 & 0.999750 & 0.000000 \\
 5 & 0.999500 & 0.999500 & 0.000000 \\
 6 & 0.999350 & 0.999350 & 0.000000 \\
 7 & 0.998900 & 0.998900 & 0.000000 \\
 8 & 0.997551 & 0.997550 & 0.000001 \\
 9 & 0.994007 & 0.994000 & 0.000007 \\
 10 & 0.986988 & 0.987050 & -0.000062 \\
 11 & 0.975836 & 0.975850 & -0.000014 \\
 12 & 0.954670 & 0.954500 & 0.000170 \\
 13 & 0.910886 & 0.910800 & 0.000086 \\
 14 & 0.826067 & 0.825350 & 0.000717 \\
 15 & 0.681002 & 0.679800 & 0.001202 \\
 16 & 0.460932 & 0.463650 & -0.002718 \\
 17 & 0.213367 & 0.212600 & 0.000767 \\
 18 & 0.046814 & 0.046650 & 0.000164 \\
 19 & 0.002129 & 0.001350 & 0.000779 \\
\hline
\end{tabular}
\caption{
 Testing the independence between $\overline{{\cal L}}_k$
and $\overline{{\cal L}}_{k+1}$ in turbo graphs
with chain length $n=64000$.
}
\label{tab:turbo}
\end{center}
\end{table}

\begin{table}
\begin{center}
\begin{tabular}{|r|r|r|r|}
\hline
 k & $P(\overline{{\cal L}}_{2k})P(\overline{{\cal L}}_{2k+2})$
 & $P(\overline{{\cal L}}_k, \overline{{\cal L}}_{k+1})$  & Difference \\
\hline
 1 & 0.999542 & 0.999542 & 0.000000 \\
 2 & 0.995460 & 0.995458 & 0.000002 \\
 3 & 0.963715 & 0.963708 & 0.000007 \\
 4 & 0.746716 & 0.746771 & -0.000055 \\
 5 & 0.097712 & 0.097333 & 0.000379 \\
\hline
\end{tabular}
\caption{
 Testing the independence between $\overline{{\cal L}}_k$
and $\overline{{\cal L}}_{k+1}$ in LDPC graphs
with $n=63000,d_v=3,d_c=5$.
}
\label{tab:ldpc}
\end{center}
\end{table}

Table \ref{tab:turbo} provides the comparison for turbo graphs for
$n=64000$.
The products of
the marginal probabilities are quite close to the joint probabilities,
indicating that the independence assumption leads
to a good approximation for turbo graphs.  Table
\ref{tab:ldpc} gives a similar results for LDPC, i.e.,
the independence assumption appears quite accurate here
also.  Thus,
 we conclude that the first independence assumption (that
the non-occurrence of cycles of length $k$ is independent
of the non-occurrence of cycles of length $k-1$ of less)
appears to be quite accurate for both turbo graphs and LDPC graphs.

Since assumption 2 is the only other approximation
being made in the analysis of the LDPC graphs, we can
conclude that it is this approximation which is less
accurate (given that the approximation and simulation
do not agree so closely overall for LDPC graphs).
Recall that the second approximation is of the form:
\begin{eqnarray}
  P(\overline{{\cal L}}_k)
 & = & P( \overline{pic}_1, \overline{pic}_2,  \ldots) \nonumber \\
 & = & \prod_i P( \overline{pic}_i \mid \overline{pic}_{i-1}, \ldots, \overline{
pic}_{1}) \nonumber \\
 & \approx & \prod_{i} P(\overline{pic}_i) \nonumber
\label{main2eqn}
\end{eqnarray}
This assumption can fail for example
when two pictures
have the first few edges in common. If one fails to be embedded on one
of these common edges, then the other will fail too.  So the best we can
hope from this approximation is that because there are so many pictures,
these dependence effects will cancel out.  In other words,
we know that
\[
 P( \overline{pic}_i) \neq
P( \overline{pic}_i \mid \overline{pic}_{i-1}, \ldots, \overline{pic}_{1})
\]
but we hope that
\[
 P( \overline{pic}_1, \overline{pic}_2,  \ldots)
\approx \prod_{i} P(\overline{pic}_i).
\]


One possible reason for the difference
between the LDPC case and the
turbo case  is as follows.
For turbo graphs,
in the expression for
the probability of embedding a picture,
\[
P_n(k,m) \approx
\frac{1}{n-\frac{m}{2}}
  \prod_{s=0}^{s=\frac{m}{2}}
    \left[  \left(1-\frac{s}{n-s}\right)
      \left(1-\frac{1}{n-2s)}\right)
    \right]^2
\]
the term $\frac{1}{n-\frac{m}{2}}$ is the most important, i.e.,
all other terms
are nearly 1. So even if two pictures share many common edges and
become dependent, as long as they do not share that most important
edge, they can be regarded as effectively independent.

In contrast, for LDPC graphs,
 the contribution from the individual edges to the total probability
tends to be more ``evenly distributed." Each edge contributes a $
\left( 1-\frac{1}{d_c} \right)$ term or a $\left( 1-\frac{1}{d_v} \right)$
term. No single edge dominates the right hand
side of
\[
P_{embed}(k=2m) =
\frac{1}{n-1}
\left( 1-\frac{1}{d_c} \right) ^{m}
\left( 1-\frac{1}{d_v} \right) ^{m-1}
\prod_{i=0}^{m-2}
\left[
        \left( 1-\frac{i}{n-1} \right)
        \left( 1-\frac{i}{c-1} \right)
\right],
\]
and, thus, the ``effective independence" may not hold as
in the case of turbo graphs.


\section{Connections to Iterative Decoding}
\label{sec:discussion}

For turbo graphs we have shown that randomly chosen nodes   are
relatively rarely on a cycle  of length 10 or less, but are highly
likely to be on a cycle of length 20 or less (for a block
length of 64000). It is interesting to conjecture about what this
may tell us about the accuracy of the iterative message-passing
algorithm in this context.

It is possible to show that there is a well-defined ``distance
effect" in message propagation for typical ADG models [17]. 
Consider a simple model where there is a hidden
Markov chain
consisting of binary-valued ${S_i}$ state nodes, $1 \le i \le N$.
In addition
there is are observed $Y_i$, one
for each state $S_i$ and
which only depend directly on each state $S_i$.
 $p(Y_i|S_i)$ is a conditional Gaussian with mean $S_i$
and standard deviation $\sigma$.
One can calculate the effect of any observed $Y_i$
on any hidden node $S_j$, $j > i$, in terms of the expected
difference between $p(S_j|Y_{j},\ldots,Y_{i+1})$
and $p(S_j|Y_{j},\ldots,Y_{i})$, 
averaged across many observations of the $Y$'s.
This average change in probability, from knowing $Y_i$,
can be shown to be proportional to $e^{-|i-j|}$, i.e.,
the effect of one variable on another dies off exponentially
as a function of distance along the chain.
Furthermore, one can show that as
the channel becomes more reliable ($\sigma$ decreases), 
the dominance of local information over information further
away becomes stronger, i.e., $Y_i$ has less effect on the
posterior probability of $S_j$ on average.

The exponential decay of information during message propagation
suggests that there may exist graphs with cycles
where the information
being propagated by a message-passing algorithm (using the 
completely parallel, or concurrent, 
version of the algorithm) can effectively ``die
out" before causing the algorithm to double count. Of course, as
we have seen in this paper, there is a non-zero probability of
cycles of length $k \ge 4$ for realistic turbo graphs, so that this
line of argument is insufficient on its own to explain the
apparent accuracy of iterative decoding algorithms.

It is also of interest to note that
that iterative decoding has been empirically observed
 to converge to stable bit decisions within 10 or so.
As shown experimentally in
[5], even beyond 10 iterations of message-passing there are still
a small fraction of nodes which typically change bit decisions.
Combined with the results on cycle length distributions in this
paper, this would suggest that it is certainly possible that
double-counting is occurring at such nodes. It may be possible to
show, however, that any such double-counting  has relatively
minimal effect on the overall quality of the posterior bit
decisions. 


\section{Conclusions}
\label{sec:conclusion} The distributions of cycle lengths in
turbo code graphs and LDPC graphs were analyzed and simulated.
Short cycles  (e.g., of length $k \le 8$) occur with relatively
low probability at any randomly chosen node. As the cycle length
increases, there is a threshold effect and the probability of a
cycle of length $k$ or less approaches 1  (e.g., for $k > 20)$.
 For turbo codes, as the block length $n$ becomes large, the
probability that a cycle of length $k$ or less exists at any
randomly chosen node behaves approximately as
$e^{-\frac{2^{k-1}-4}{n}}, k \ge 4$.  The S-random construction is
shown to eliminate very short cycles and for larger cycles
results
in only a small systematic decrease in the probability of such cycles.
For LDPC codes the analytic approximations are less accurate than for the
turbo case (when compared to simulation results). Nonetheless
the distribution as a function of $k$ shows qualitatively similar
behavior to the distribution for turbo codes, as a function of
cycle length $k$.
In summary,
the results in this paper demonstrate that the cycle lengths in turbo
graphs and LDPC graphs have a specific distributional character.
We hope that this information can be used to further understand
the workings of iterative decoding.

\section*{Acknowledgments}
The authors are grateful to R. J. McEliece and the coding group at
Caltech for many useful discussions and feedback.

\section*{References}

\begin{enumerate}
\item
C. Berrou, A. Glavieux, and P. Thitimajshima (1993).
Near Shannon limit error-correcting coding and decoding: Turbo codes.
{\it Proceedings of the IEEE International Conference on Communications}.
pp. 1064-1070.

\item
R.J. McEliece, D.J.C. MacKay, and J.-F. Cheng (1998).
Turbo Decoding as an Instance of Pearl's `Belief Propagation' Algorithm.
{\it IEEE Journal on Selected Areas in Communications,} SAC-16(2):140-152.

\item
F. R. Kschischang, B. J. Frey (1998).
Iterative Decoding of Compound Codes by Probability Propagation
in Graphical Models.
{\it IEEE Journal on Selected Areas in Communications,} SAC-16(2):219-230.

\item
P. Smyth,  D. Heckerman, and M. I. Jordan (1997).
`Probabilistic independence networks for hidden Markov probability
models,' {\it Neural Computation}, 9(2),  227--269.

\item
B. J. Frey (1998).
{\it Graphical Models for Machine Learning and Digital Communication.}
MIT Press: Cambridge, MA.

\item
J. Pearl (1988),
{\it Probabilistic Reasoning in Intelligent Systems:
Networks of Plausible Inference.}
Morgan Kaufmann Publishers, Inc., San Mateo, CA.

\item L. R. Bahl, J. Cocke, F. Jelinek, and J. Raviv (1974).
`Optimal decoding of linear codes for minimizing symbol error rate,'
{\it IEEE Transactions on Information Theory}, 20:284--287.

\item Y. Weiss (1998).
`Correctness of local probability propagation in graphical models with loops,' submitted to
{\it Neural Computation}.

\item R. J. McEliece, E. Rodemich, and J. F. Cheng,
`The Turbo-decision Algorithm,'
in {\it Proc. Allerton Conf. on Comm., Control, Comp.}, 1995.

\item
S. Dolinar and D. Divsalar (1995),
{\it Weight Distributions for Turbo Codes
Using Random and Nonrandom Permutations.}
TDA Progress Report 42-121 (August 1995),
Jet Propulsion Laboratory, Pasadena, California.

\item
R. G. Gallager (1963),
{\it Low-Density Parity-Check Codes.}
Cambridge, Massachusetts: MIT Press.

\item
D.J.C. MacKay, R.M. Neal (1996),
{\it Near Shannon Limit Performance of
Low Density Parity Check Codes},
published in {\it Electronics Letters},
also available from {\tt http://wol.ra.phy.cam.ac.uk/}.

\item
K. S. Andrews, C. Heegard, and D. Kozen, (1998).
  `Interleaver Design Methods for Turbo Codes,'
  {\it
Proceedings of the 1998 International Symposium on Information Theory},
 pg.420.

\item
C. Heegard and S. B. Wicker (1998),
 {\it Turbo Coding},
  Boston, MA: Kluwer Academic Publishers.

\item
T. Richardson, R. Urbanke (1998),
{\it The Capacity of Low-Density Parity Check Codes
under Message-Passing Decoding},
preprint, available at \\
{\it http://cm.bell-labs.com/who/tjr/pub.html}.

\item
M. G. Luby, M. Mitzenmacher, M. A. Shokrollahi, D. A. Spielman (1998),
`Analysis of Low Density Codes and Improved Designs Using Irregular Graphs,'
      in {\it Proceedings of the 30th ACM STOC}. Also available
online at \\
 {\it http://www.icsi.berkeley.edu/~luby/index.html}.

\item
X. Ge and P. Smyth,
`{\it Distance effects in message propagation}',
in preparation.

\end{enumerate}

\end{document}